\documentclass[11pt]{article}
\usepackage{moriond1,epsfig}

\def\be{\begin{equation}}
\def\ee{\end{equation}}
\def\bea{\begin{eqnarray}}
\def\eea{\end{eqnarray}}

\def\lsim{\mathrel{\rlap{\lower4pt\hbox{\hskip1pt$\sim$}}
    \raise1pt\hbox{$<$}}}                % less than or approx. symbol
\def\gsim{\mathrel{\rlap{\lower4pt\hbox{\hskip1pt$\sim$}}
    \raise1pt\hbox{$>$}}}                % greater than or approx. symbol
\newcommand{\qsq}{\ensuremath{Q^2}}
\newcommand{\qsqw}{Q^2}
\newcommand{\jpsi}{$J/\psi$}

\newcommand{\wgp}{$W_{\gamma p}$}
\newcommand{\wgpw}{W_{\gamma p}} 
\newcommand{\GeV}{\,\mbox{GeV}}

\newcommand{\GeVt}{\,\mbox{\GeV$^2$}}
\newcommand{\gevt}{\,\mbox{\GeV$^2$}}
\newcommand{\gev}{\,\mbox{GeV}}
\newcommand{\gmt}{\,\mbox{\GeV$^{-2}$}}
\begin{document}
\vspace*{4cm}
\title{New Results on Vector Meson Production at HERA}

\author{ B. NAROSKA,\\on behalf of the H1 and ZEUS collaborations}% \today,\the\time}

\address{University of Hamburg, II. Inst. f. Experimentalphysik\\
Luruper Chaussee 149, 22600 Hamburg, Germany\\e-mail:naroska@mail.desy.de}

\maketitle\abstracts{
New results on elastic (exclusive) production of vector mesons are 
presented, mainly on $J/\psi$ and $\phi$  mesons with emphasis on an
interpretation of the data within pQCD.}

\section{Introduction}

Production of vector mesons at HERA has become a rich field of
experimental and theoretical research. In this report only fairly new
results will be shown concentrating on $J/\psi$ and $\phi$ meson. For an 
overview of data on $\rho$ mesons
see~\cite{rho}. Vector mesons can be produced elastically (i.e. exlusively), 
i.e. in
the reaction $ep \rightarrow eVp$, where $V$ denotes the vector meson and 
the proton remains a proton.
The largest background is the proton dissociative process where the proton
breaks up into a low mass system. H1 and ZEUS have developed efficient 
procedures to remove and correct for this and other backgrounds~\cite{jpsi1,jpsi2,jpsi3}.

\begin{table}[b!]
\caption{Kinematical Quantities.\label{tab:kin}}
%\vspace{-0.4cm}
\begin{center}
\begin{tabular}{lll}
\hline
%& &  \\
$ep$ center of mass energy squared &  $s = (p+k)^2$ & $k$,$p$ incoming lepton, proton\\
neg. momentum transfer squared & $Q^2 = -q^2= -(k-k^\prime)^2$  &  $k^\prime$ scattered lepton\\
scaled energy transfer & $y=(p\cdot q) / (p \cdot k) $  &  $q$ exchanged photon\\
$\gamma\,p$ center of mass energy squared & $\wgpw^2 = (p+q)^2=ys-Q^2$  \\
%& &   \\ 
momentum transfer proton   & $t = (p-p^\prime)^2$  &  $p^\prime$ scattered proton \\
%&  \\
\hline
\end{tabular}
\end{center}
\end{table}

There is by now no doubt that pQCD can describe the elastic production of 
vector mesons~\cite{fks,teubner} via exchange of a colour neutral system of
gluons, i.e. in leading order via exchange of two gluons. 
The QCD scale can be given
by the mass of the vector meson, as in the case of the $J/\psi$ or by $Q^2$
as for the ``light'' vector mesons $\rho, \omega$ and $\phi$. 
Whether the momentum transfer
 $t$ between incoming and outgoing proton (or the outgoing dissociated
proton), can also serve as a hard scale is still under experimental 
investigation~\cite{zphi3}.
A signature for the ``hard'' behaviour is the fast rise of the integrated 
$\gamma\,p$ cross section for vector meson production with \wgp.

\section{Total Cross Sections for \boldmath$J/\psi$ and \boldmath$\phi$ Mesons}

\begin{figure}[t!]
\setlength{\unitlength}{1cm}
\begin{picture}(16.,6.)
\put(0.5,-0.2){\epsfig{file=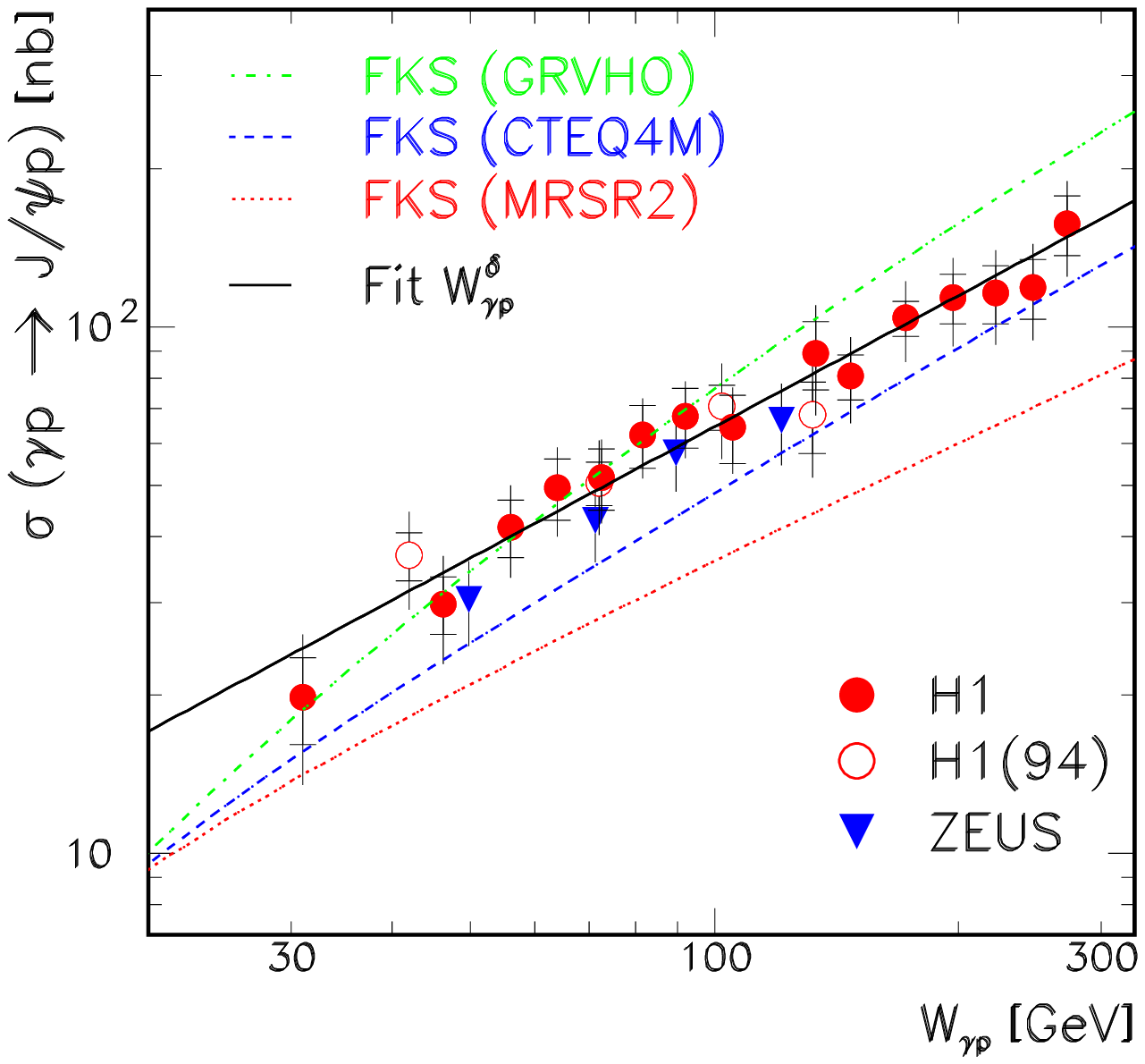,height=6.5cm}}
\put(8.,0.5){\epsfig{file=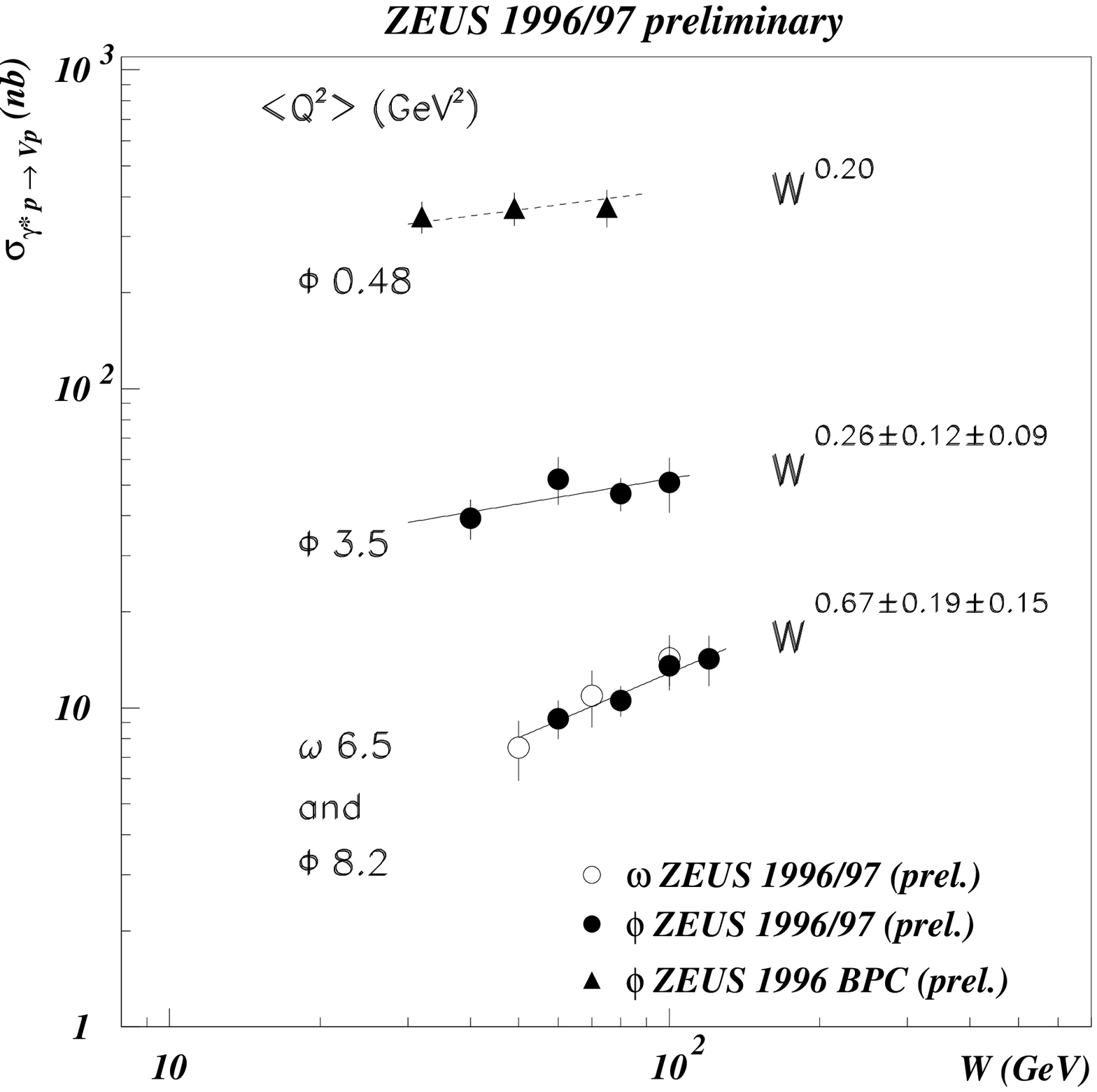,bbllx=0,bblly=70,bburx=525,bbury=540,width=6.5cm,clip=}}
\end{picture}
\caption{Total cross sections for $\gamma p \rightarrow V\,p$ as a function 
of $W_{\gamma p}$ for $V=J/\psi$  (left) and for $V=\phi$ (right). 
Fits to the data of the form  $W_{\gamma p}^\delta$ are shown. In the left 
plot also results 
from pQCD calculations are given using various gluon density distributions.
\label{fig1}}
\end{figure}

Photoproduction of \jpsi\ mesons has been measured by the H1 
collaboration~\cite{jpsi3} in the range of $26\leq\wgpw \leq 285\,\gev$. 
Leptonic decays into $e^+e^-$ or $\mu^+\mu^-$ are used depending on the detector region.
The data are shown in Fig.~\ref{fig1} and are compared with results from 
pQCD calculations by Frankfurt et al.~\cite{fks} using various gluon density functions. 
The main prediction concerns the slope of the data which is seen to be well represented by the calculation using the CTEQ4M or MRSR2 gluon densities.
A fit to the data of the form $\wgpw^\delta$ yields $\delta=0.83\pm0.07$ 
which is much larger than $\delta\sim 0.3$ found in soft processes.

New data on the production of $\phi$ mesons at $Q^2 \gsim 1\,\GeVt$ became
available from H1~\cite{zphi1} and ZEUS~\cite{zphi3,zphi2} both using the decay to $K^+K^-$. ZEUS also shows first results on $\omega$ production in a range $3<\qsq<20\,\gevt$
using the decay $\omega\rightarrow \pi^+\pi^-\pi^0$.
The integrated  cross sections $\sigma(\gamma^* p
\rightarrow \phi p)$  and  $\sigma(\gamma^* p
\rightarrow \omega p)$ are shown in Fig. \ref{fig1}. Parameterising them as 
$\wgpw^\delta$ a clear increase
of $\delta$ with $Q^2$ can be seen as has been found for the $\rho$ meson~\cite{rho} .

We conclude: if the fast increase of $\sigma(\gamma p \rightarrow Vp)$ with
\wgp\ is indeed a signature for a hard process, either $M^2_V$ or $Q^2$ serve
as hard scales. %This has been proposed many times in the literature.

\section{Determination of the Regge Trajectory for \boldmath$J/\psi$ Photoproduction}

In order to determine the properties of the exchange mediating the
interaction between the \jpsi\ meson and the proton, Regge language
is used. The Regge trajectory is determined by H1 for photoproduction of 
$J/\psi$ mesons assuming a simple linear form 
$\alpha(t) = \alpha_0 + \alpha' t$.
$\alpha(t)$ determines the dependence of the cross sections on the energy 
\wgp\ as  $W_{\gamma p}^{4(\alpha(t) - 1)}$. In ``hard''  interactions  $\alpha'$
is expected to be small~\cite{fks} while in ``soft'' reactions 
$\alpha'\approx0.25\,\gmt$ has been found~\cite{1pom}.

\begin{figure}[p]
\psfig{figure=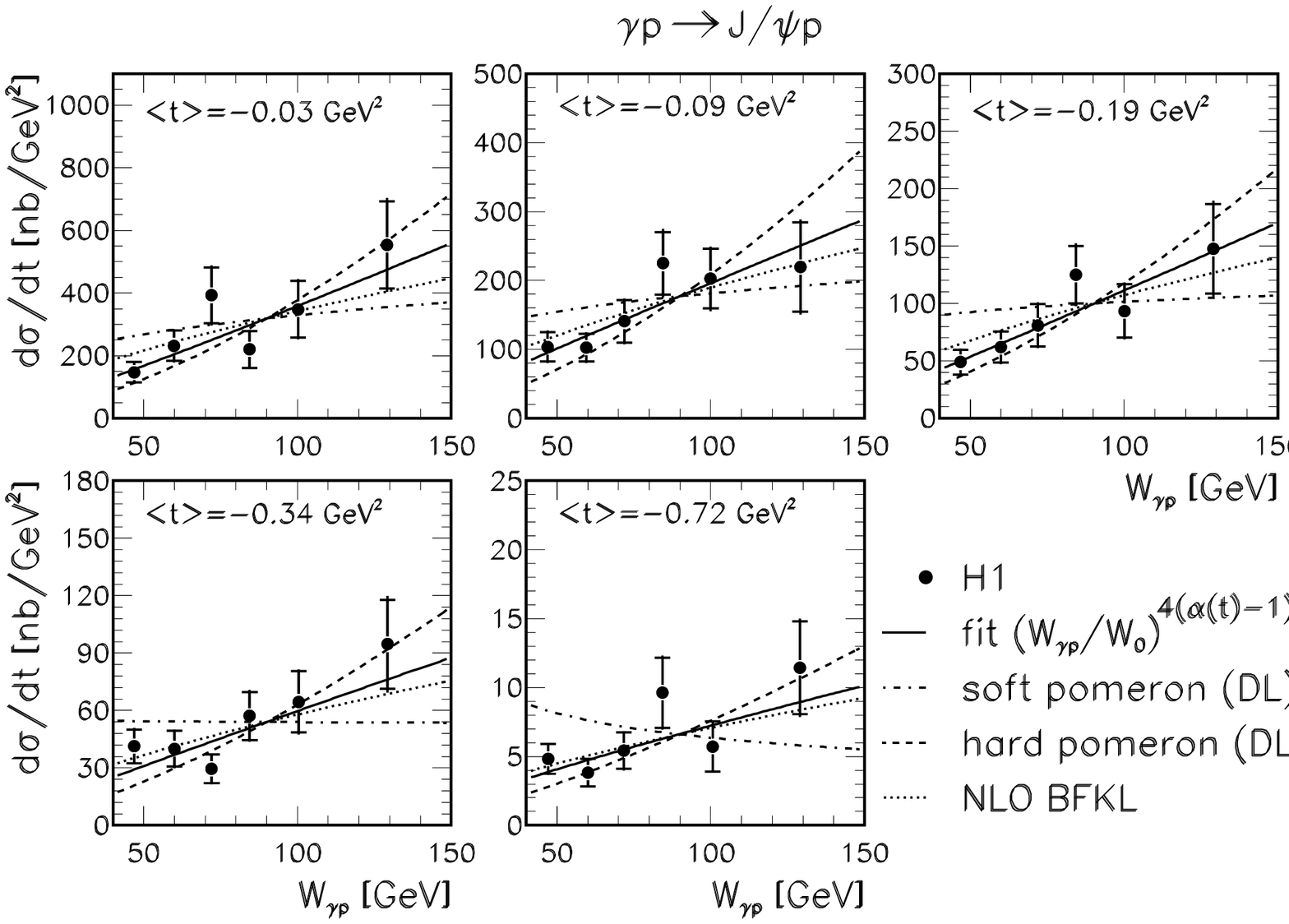,height=7cm}
\psfig{figure=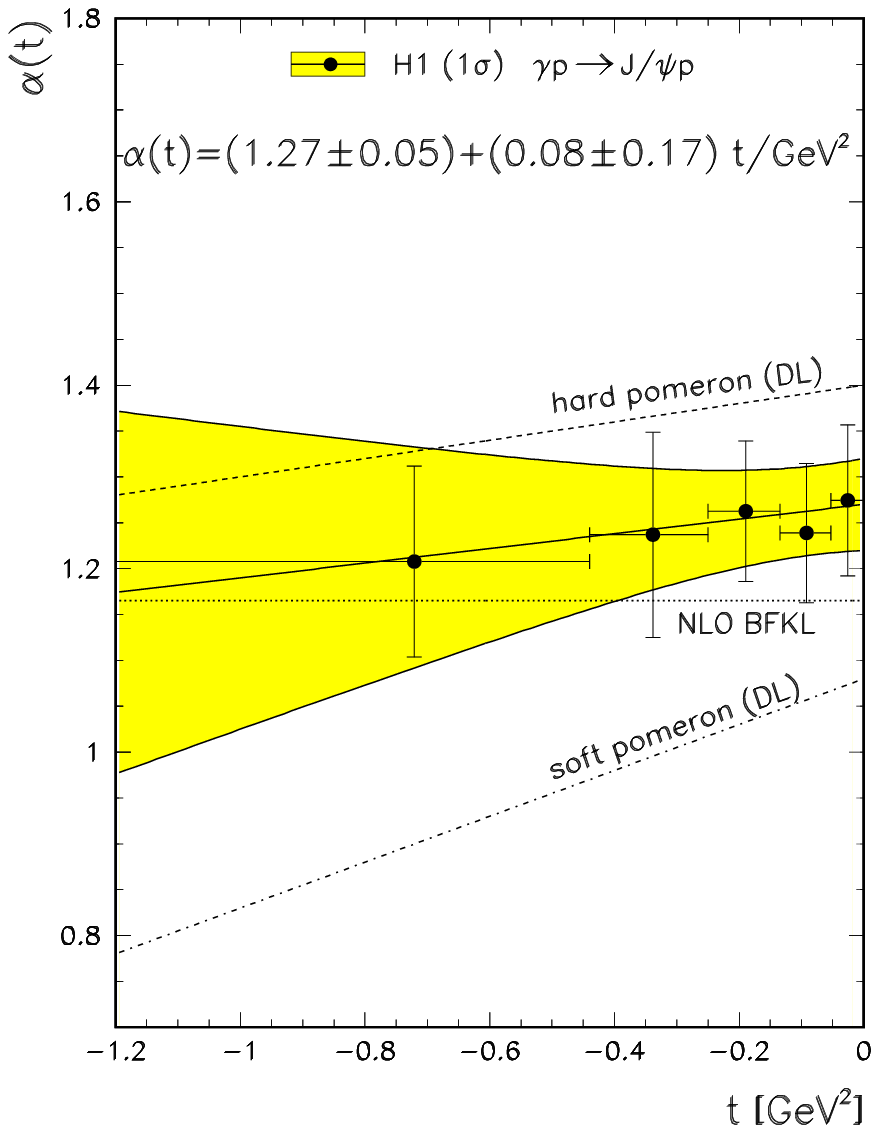,height=7cm}
\caption{Left: The differential cross section $d\sigma/dt$ as a function of 
           $W_{\gamma p}$ in five bins of $t$ together with a fit of the form 
           $d\sigma/dt=N\cdot(W_{\gamma p}/W_0)^{4(\alpha(t)-1)}$ (solid line). 
           Predictions of models using soft 
           and hard pomeron trajectories  
           are shown. 
 Right: The measured Regge trajectory %\protect{$\alpha(t)=\alpha_0+\alprim t$} 
           for the process $\gamma p \rightarrow J/\psi\,p$. The solid line shows the 
           result of the fit. The one standard deviation contour is indicated by a 
           shaded band. Also shown are the soft and the hard Donnachie-Landshoff 
           pomeron trajectories~\protect\cite{1pom} and a result based on a NLO BFKL 
           calculation~\protect\cite{bfklpom}.
\label{fig2}}
%\end{figure}

%\begin{figure}
\setlength{\unitlength}{1cm}
\begin{picture}(16.,12.5)
\put(7.5, 6.7){\psfig{figure=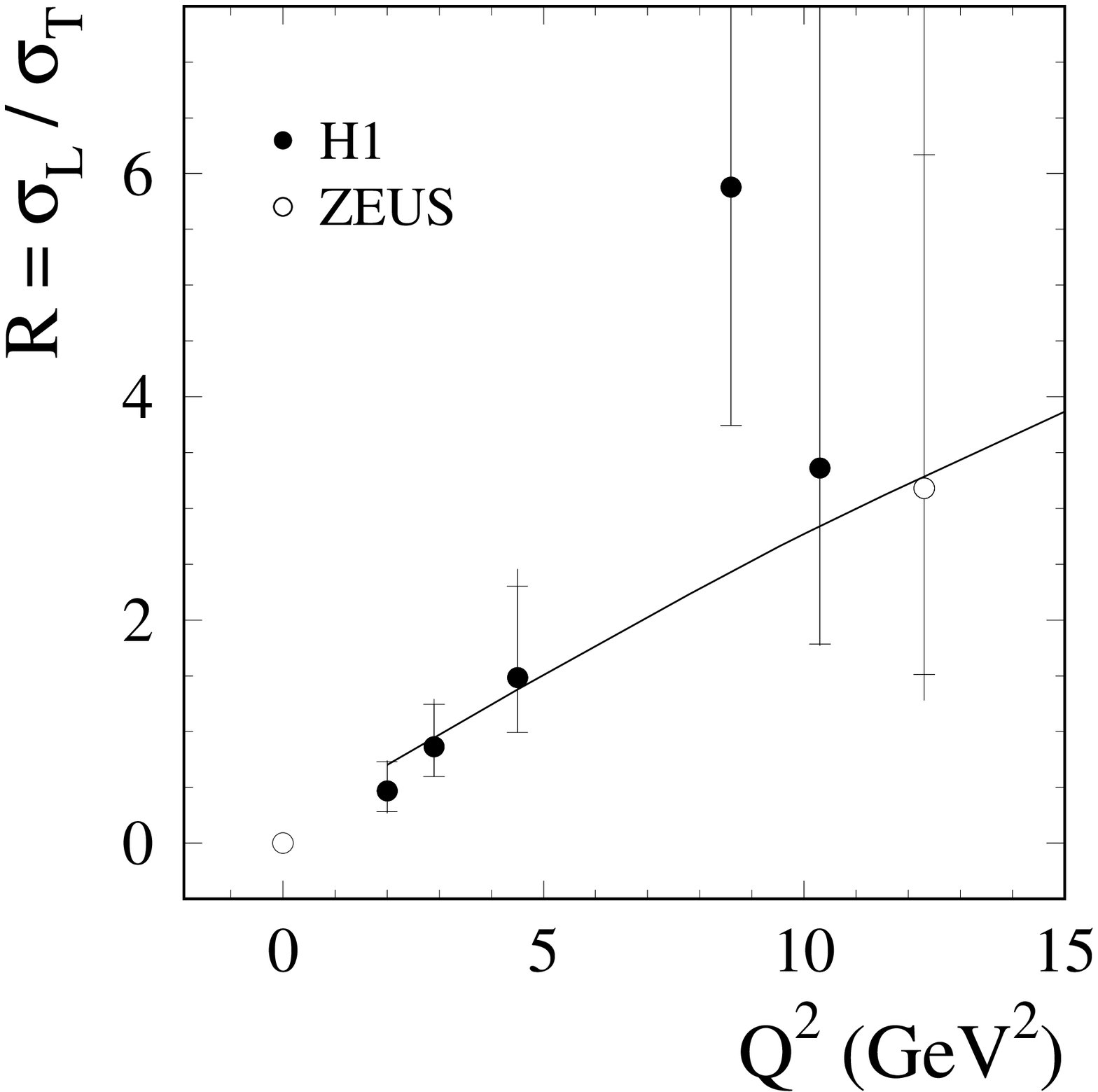,height=5.8cm}}
\put(1.,6.){\psfig{figure=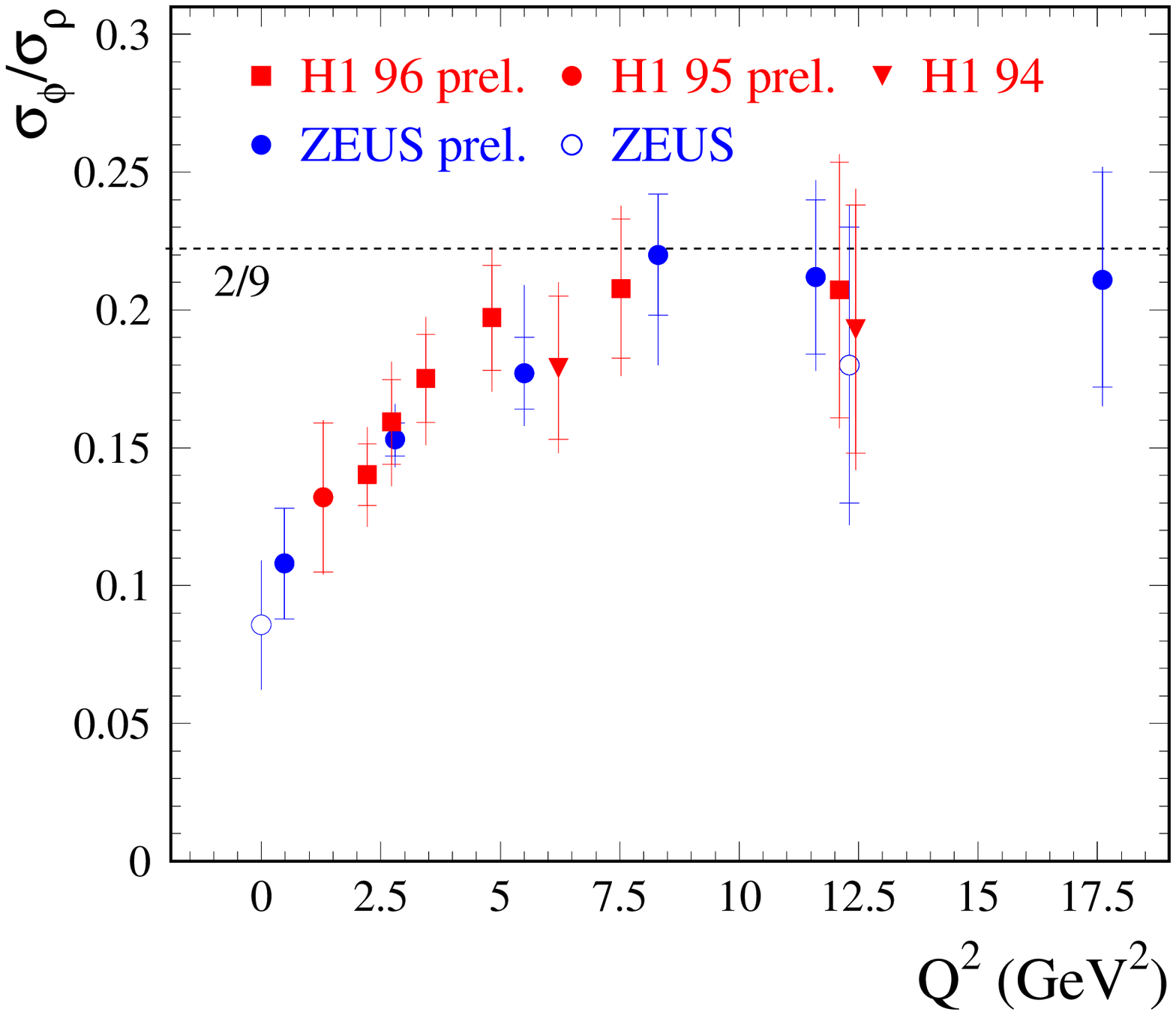,height=6.5cm}}
\put(1.,0.5){\epsfig{file=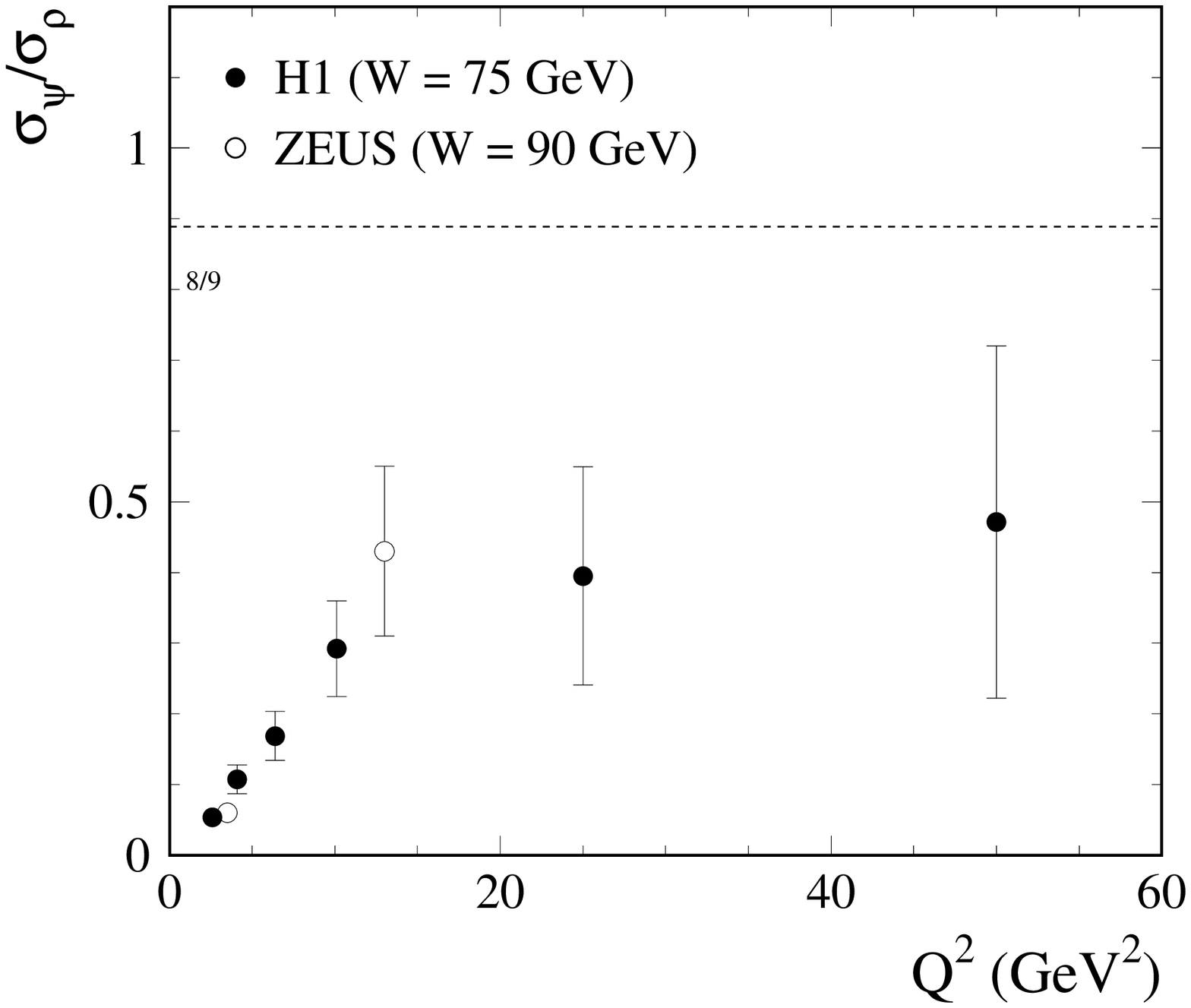,height=6.5cm,clip=}}
\put(7.5,-0.2){\psfig{figure=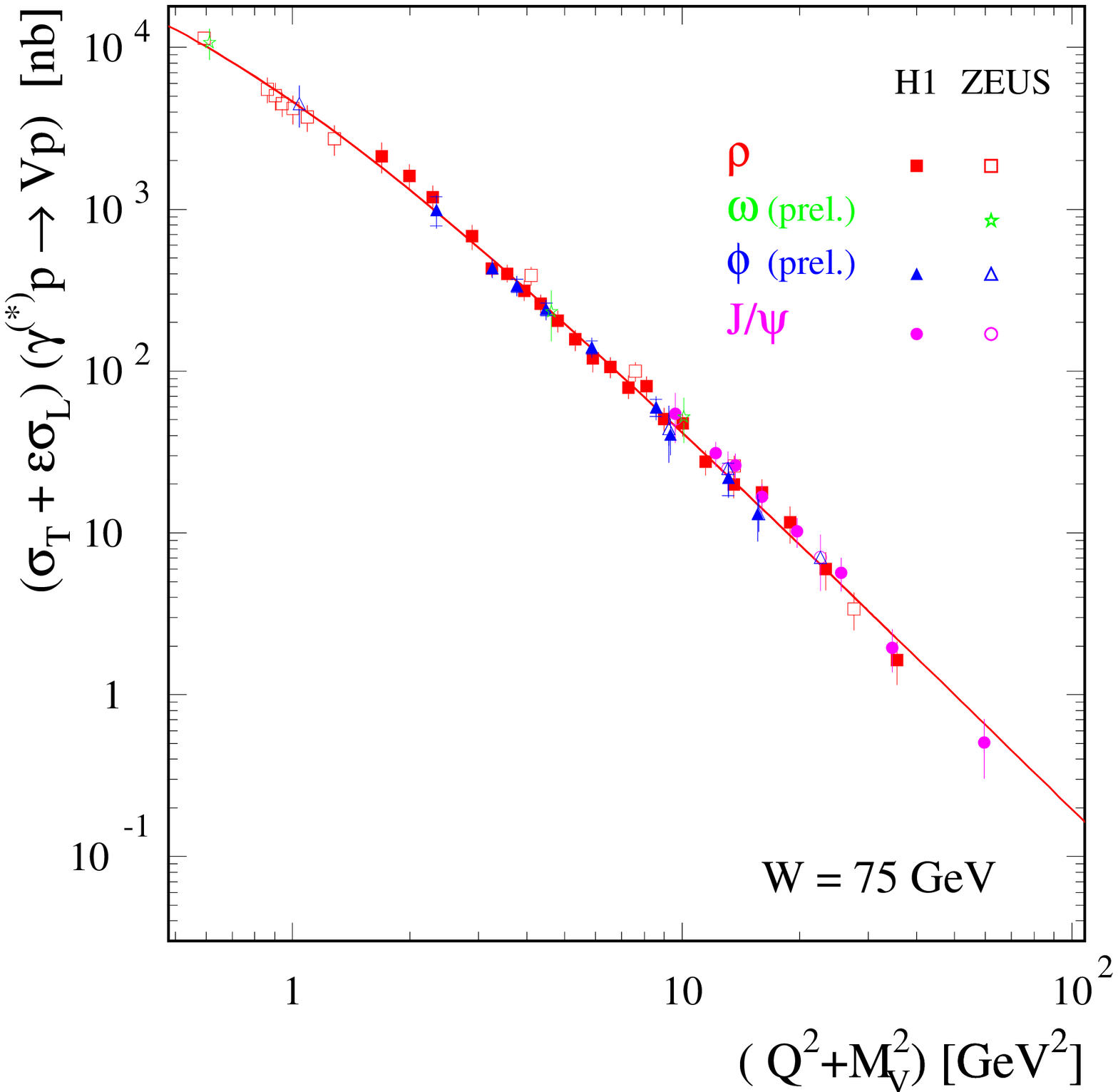,height=7.6cm}}
\put(5.5,5.3){\small 8/9}
\put(5.,2.7){b)}
\put(5.,8.){a)}
\put(9.,2.){d)}
\put(12.,8.){c)}
\end{picture}
\caption{The ratio of the integrated cross sections  (a) 
$\sigma_{\phi}/\sigma_\rho$ and (b) $\sigma_{J/\psi}/\sigma_\rho$ as a 
function of \protect\qsq.
 (c)$R=\frac{\sigma_L}{\sigma_T}$ 
for $\phi$ meson production as a function of $Q^2$. 
The curve is a result from a pQCD calculation~\protect\cite{teubner}.
(d) The integrated cross sections for vector mesons scaled by the SU(4) 
ratios as a function of $\qsqw+M_V^2$. The data are scaled to 
\protect{$\wgpw=75\,\gev$}.
\label{fig3}}
\end{figure}

H1 determined the trajectory for $J/\psi$ production using data from one
experiment only and thus avoiding normalisation problems between data from
different experiments.
The data in the range $40 \leq W_{\gamma p} \leq 150$ GeV are used. 
$d\sigma/dt$ is determined at 5 values of $t$. For these 5 values of $t$ the 
dependence of $d\sigma/dt$ on $W_{\gamma p}$ is shown in 
Fig. \ref{fig2} (left).
A fit to the form $W_{\gamma p}^{4(\alpha(t) - 1)}$ is performed and the 
resulting values 
for $\alpha(t)$ are shown in Fig. \ref{fig2} (right). A linear fit to $\alpha(t)$ 
yields the parameters of the fitted
trajectory which are also given in the figure. The 1$\sigma$ contour 
is shown as are results of various predictions for the pomeron based on non-perturbative models and on a NLO BFKL calculation~\cite{bfklpom}. 
In contrast to the results for the $\rho$ 
and $\phi$ mesons $\alpha_0$ for the $J/\psi$ is larger and $\alpha'$ is 
compatible with 0.

\section{The Scale for Production of Vector Mesons}

When $Q^2$ increases the exchanged photon acquires a longitudinal
component. $R = \sigma_L/\sigma_T$ has previously been measured for the
$\rho$ and $J/\psi$ meson~\cite{rho,jpsi2}. In Fig. \ref{fig3}c the ratio $R$ is shown 
for the $\phi$ meson as a function of \qsq\ and is seen to increase. 
The curve based on pQCD and 
parton hadron duality~\cite{teubner} gives a good description of the data.

If in elastic vector meson production the photon couples indeed to 
quarks (as opposed to coupling to a hadron) and the
interaction of the quark pair of the vector meson with the proton is
universal the ratio of cross sections should only depend on the quark content 
of the vector mesons. This is indeed observed in Fig. \ref{fig3}a,b where 
$\sigma_\phi/\sigma_\rho$ and $\sigma_{J/\psi}/\sigma_\rho$ are plotted 
as functions of \qsq. The ratio is seen 
to approach the value of $2/9$ for the $\phi$ 
expected from the simple SU(4) quark counting. The ratio 
$\sigma_{J/\psi}/\sigma_\rho$ increases slowly and is still below the 
expected value of $8/9$ at $\qsqw\sim40\,\gevt$, 
but the errors are still large.
 A universal behaviour of the cross section for all vector mesons 
is observed  in Fig. \ref{fig3}d, where all available data,
for $\rho, \omega, \phi$ and $J/\psi$ mesons are shown as a function
of $Q^2 + M^2_V$. The data have been scaled to a common
$W_{\gamma p} = 75$ GeV using the measured $W_{\gamma p}$ dependence.
They have been scaled by the quark charges assuming the SU(4) ratio
$\rho\ :\ \omega\ :\ \phi\ :\ J/\psi = 9\ :\ 1\ :\ 2\ :\ 8$.
The data are seen to agree well with each other and can be described by 
a function $(Q^2 + M^2_V + a)^b$ with $a=0.42\pm0.09\,\gevt$ and $b=-2.37\pm0.10$, 
which was fitted to the $\rho$ data. 
 We can conclude that within present errors  $(Q^2 + M^2_V)$ is a good scale for 
photoproduction of vector mesons at low values of $t$.

\section*{Acknowledgments}
I thank the organisers for an interesting meeting and my H1 and ZEUS colleagues for discussions. Skiing after 10 years was fun!

\end{document}